# Graphene Helicoid: The Distinct Properties Promote Application of Graphene Related Materials in Thermal Management

*Haifei Zhan[1,2], Gang Zhang[3,\*], Chunhui Yang[1], and Yuantong Gu[2,\*\*]*

*Corresponding author. E-mail: zhangg@ihpc.a-star.edu.sg (Gang Zhang)*
*\*\*Corresponding author. E-mail: yuantong.gu@qut.edu.au (Yuantong Gu)*

[1]School of Computing, Engineering and Mathematics, Western Sydney University, Locked Bag 1797, Penrith NSW 2751, Australia

[2]School of Chemistry, Physics and Mechanical Engineering, Queensland University of Technology (QUT), Brisbane QLD 4001, Australia

[3]Institute of High Performance Computing, Agency for Science, Technology and Research, 1 Fusionopolis Way, Singapore 138632, Singapore

**ABSTRACT:** The extremely high thermal conductivity of graphene has received great attention both in experiments and calculations. Obviously, new feature in thermal properties is of primary importance for application of graphene-based materials in thermal management in nanoscale. Here, we studied the thermal conductivity of graphene helicoid, a newly reported graphene-related nanostructure, using molecular dynamics simulation. Interestingly, in contrast to the converged cross-plane thermal conductivity in multi-layer graphene, axial thermal conductivity of graphene helicoid keeps increasing with thickness with a power law scaling relationship, which is a consequence of the divergent in-plane thermal conductivity of two-dimensional graphene. Moreover, the large overlap between adjacent layers in graphene helicoid also promotes higher thermal conductivity than multi-layer graphene. Furthermore, in the small strain regime (< 10%), compressive strain can effectively increase the thermal conductivity of graphene helicoid, while in the ultra large strain regime (~100% to 500%), tensile strain does not decrease the heat current, unlike that in generic solid-state materials. Our results reveal that the divergence in thermal conductivity, associated with the anomalous strain dependence and the unique structural flexibility, make graphene helicoid a new platform for studying fascinating phenomena of key relevance to the scientific understanding and technological applications of graphene-related materials.

## 1. Introduction

Continuing miniaturization of electronic devices requires efficient heat management due to the significantly increased power density. It has been a great enthusiasm to seek materials with excellent heat conductivity for next generation of integrated circuits or strongly suppressed thermal conductivity for thermoelectric energy conversion devices. In terms of thermal management, the carbon-based materials occupy a unique place as their room-temperature thermal conductivity spans over fiver orders of magnitude from ~ 0.01 W/mK in amorphous carbons to thousands W/mK in diamond or graphene.[1] For instance, the thermal conductivity of carbon nanotubes (CNTs) is measured as high as ~3,000 to 3,500 W/mK at room-temperature.[2] Experimental measurements show that the suspended single-layer graphene has a higher thermal conductivity of ~ 2500 - 5300 W/mK at room-temperature.[1, 3-5] The graphene filler is able to significantly enhance the thermal conductivity of nanocomposites that are used as thermal interface materials.[6-7]

Although ultra-high in-plane thermal conductivity is observed in monolayer graphene, its application in thermal management is limited by its atomic thin cross-sectional area (as schematically shown in **Figure 1**a). Following Fourier's law, the total heat energy transferred by a heat channel can be calculated as $Q = -S \times \kappa \times \nabla T$, where $S$ is the cross-sectional area, $\kappa$ is thermal conductivity, and $\nabla T$ is the temperature gradient. As the thickness of monolayer graphene is only 3.4 Å, the total heat dissipation rate by the in-plane thermal conduction through individual monolayer graphene is still limited. Moreover, in many integrated circuit (IC) devices, a major part of the heat generated may dissipate in the through-plane direction.[8-9] Thus the multi-layer graphene (MLG)-based thermal management strategy is an ideal alternative as shown in Figure 1b. Unfortunately, in contrast to its high in-plane thermal conductivity, the cross-plane thermal conductivity of multi-layer graphene is more than 3 orders of magnitude smaller.[1] Earlier study reported that the cross-plane thermal conductivity of MLG is only around ~ 0.7 W/mK at room temperature,[10] because the weak van der Waals (vdW) interlayer interaction dominates the cross-plane thermal conduction.

The intrinsically low cross-plane thermal conductivity of MLG has attracted extensive studies on thermal conductance modulation. Previous studies have attempted to integrate the thermal transport characteristics of single layer graphene into the MLG by using the folded nanoribbon structure to mimic the multi-layered structure. It is found that the folded structure can be used to effectely tune the thermal conductance of the layered structure.[11-12] Besides the simple pile of monolayers, another direct inspiration to assemble layered structure is enlightened by the double helical structure of DNT or the twisted shape of polyermer lamellae.[13] Considering the GNR as elastic strips, a rich range of layered configurations can be constructed with minima energy by varying the hyperbolic two-dimensional reference matrics.[14] An ideal layered structure under this concept is a helicoid structure (see Figure 1c), which is analogue to a hollow MLG structure with a hole along its thickness direction (Figure 1b). Analogous helicoid structures have been widely observed in synthesized nanomaterials, biological and self-assembled organic systems, such as the helical ZnO nanobelts,[15] twisting carbon nanobibbons,[16] and $MoS_2$ spiral pyramid.[17] Evidently, the helicoid structure is similar to a curved GNR and has the same number of atoms (also similar mass density) compared with its multi-layer counterpart. Recent studies show that such helicoid structure could bring superior inductance by imposing electrical current,[18] and can withstand huge elastic deformations with a strain range up to 1000%.[19] Therefore, this work aims to probe how the graphene helicoid (GH) structure can be used to modulate the thermal conductivity, using large-scale molecular dynamics (MD) simulations. The effects of cross-plane strain on thermal transport properties are assessed and the underlying physical mechanisms are further discussed. Our study provides practical guidance to experiments for engineering of controllable thermal conductivity in graphene-related materials.

The GH structure was constructed according to the screw dislocations as observed abundantly in annealed pyrolytic graphite.[20-21] A representative zigzag-edged structure was chosen in this work. As shown in **Figure 1c**, the GH was constructed through a single screw dislocation of graphene nanoribbon ($b$, where $|b|$ = 3.35 Å), as constrained by three geometrical parameters, i.e., outer radius ($R$), inner radius ($r$) and turn/pitch number ($N$). The width and total height of the GH can be calculated with $w = R - r$ and $L_{tot} = N \times |b|$, respectively. The initial GH structure has an AA stacking sequence. To note that the non-periodic boundary conditions in the lateral directions allow the adjacent layers to shift in order to accommodate the lowest energy status during relaxation process. At higher temperature, the out-of-plane deflections of the adjacent shifted layers would lead to occasional formation of C-C bonds at either inner or outer edges. To avoid such a bond formation, we considered a hydrogen-terminated edge where each edge carbon atom corresponds to the molecular group CH (see Figure 1).

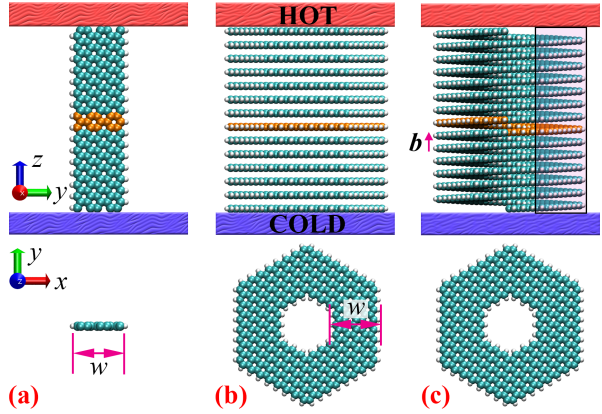

**Figure 1.** Schematic view of the heat transfer scenarios. (a) graphene nanoribbon, (b) multilayer graphene (MLG); and (c) GH constructed from screw dislocations. Upper panel is the front view, and bottom panel is the cross-sectional view. Black rectangular represents the equivalent GNR projected along the axial direction of the GH.

## 2. Computational Methods

For all MD simulations, the widely-used adaptive intermolecular reactive empirical bond order (AIREBO) potential was employed to describe the C-C and C-H atomic interactions.[22-23] This potential includes short-range interactions and long range vdW interactions, which has been shown to well represent the binding energy and elastic properties of carbon materials. It adopts Lennard-Jones term to describe the van der Waals interactions, which has been reported to reasonably capture the vdW interactions in multi-layer graphene,[24] multi-wall carbon nanotubes,[25] carbon nanotube bundles,[26] and hybrid carbon structure.[27]

Nonequilibrium molecular dynamics (NEMD) simulations were employed to acquire the heat transfer in the GH structure at 300 K. The samples were firstly optimized by the conjugate gradient minimization method and then equilibrated using Nosé-Hoover thermostat[28-29] for 400 ps. Non-periodic boundary conditions were applied in all directions. The temperatures of the heat source and sink were kept at 310 and 290 K, respectively, by using the Langevin thermostat.[30] The system was firstly simulated for 0.8 ps to arrive a steady state. After the system reached non-equilibrium steady state, the thermal conductivity was calculated within a time interval of 0.8 ns and repeated four times (over 3.2 ns). The standard deviation was estimated from these four values. A small time step of 0.2 fs was used for all calculations with all MD simulations being performed under the software package LAMMPS.[31] It is worth mentioning that the AIREBO potential is developed for local interatomic forces, not for long range interaction. Although the Lennard-Jones term is introduced to describe the interlayer vdW coupling, it may underestimate the absolute value of thermal conductivity. Note that MD simulation is a classical approach to probe the thermal behaviours of nanomaterials, which relies on the classical distribution. While it is reported that without changing temperature, it still can provide qualitatively description of external influence on the thermal conductivity,[32] for example, impurity effect[33-34] or size effect.[35]

## 3. Results and Disucssion

### 3.1 Increased thermal conductivity with thickness of GH

Firstly, we assess how the thermal transport of GH structure differs from its MLG counterpart by varying the thickness or layer number. We calculated the thermal conductivity according to $\kappa = -J/\nabla T$. Here, $\nabla T$ is the temperature gradient, and heat flux $J$ is defined as the energy injected into/removed from the heat source/sink across unit area per unit time, i.e., $J = Q/S$ with $Q$ and $S$ representing the heat current and cross-sectional area, respectively. The cross-sectional area $S$ of the GH or MLG is approximated as a hollow hexagon. In this work, one turn (or unit) at each end of the structure is fixed and the adjacent three layers at each end are grouped as heat source and sink, respectively. **Figure 2a** illustrates the temperature profile of GH at the non-equilibrium steady-state (averaged within simulation time of 2 ns), which preserves a good linear relationship between the heat source and sink. Due to the negligible temperature jump at the two ends, for simplicity, in this work we calculate the temperature gradient using constant temperature difference ($\Delta T$ = 20 K) and thickness between the heat source and sink, i.e., $\nabla T = \Delta T/\Delta L$, with $\Delta L$ as the thickness of GH/MLG between heat source and sink.

Figure 2b shows the thermal conductivity of GH and MLG in the thickness direction as a function of thickness (which are later referred as axial and cross-plane thermal conductivity to distinguish the heat transport mechanisms, respectively). As is seen, $\kappa$ of GH increases monotonically from about 0.23 ± 0.004 to ~ 0.77 ± 0.06 W/mK when the effective thickness ($\Delta L$) increases from 1.005 to 7.035 nm. On the other hand, the $\kappa$ of MLG firstly experiences a similar monotonic increase trend, and then converges to around 0.45 W/mK when the MLG thickness is over 3.685 nm (corresponding to 10 effective layer). This value is consistent with the experimentally reported cross-plane thermal conductivity of few layer graphene (FLG, with a thickness of ~ 35 nm), which is about 0.7 W/mK at room temperature.[10] Overall, the GH exhibits a higher heat transfer capability compared with MLG, and the absolute difference of $\kappa$ between GH and MLG increases when the thickness increases (see Supporting Information **S1**). It is worth mentioning that we also calculated the thermal conductivity of GHs with the same thickness but different width (ranging from ~0.4 to 1.4 nm), and found that the calculated $\kappa$ is almost width independent in the considered range of width (see Supporting Information **S1**). In the following sections, we fix the width of GH and MLG as ~0.92 nm.

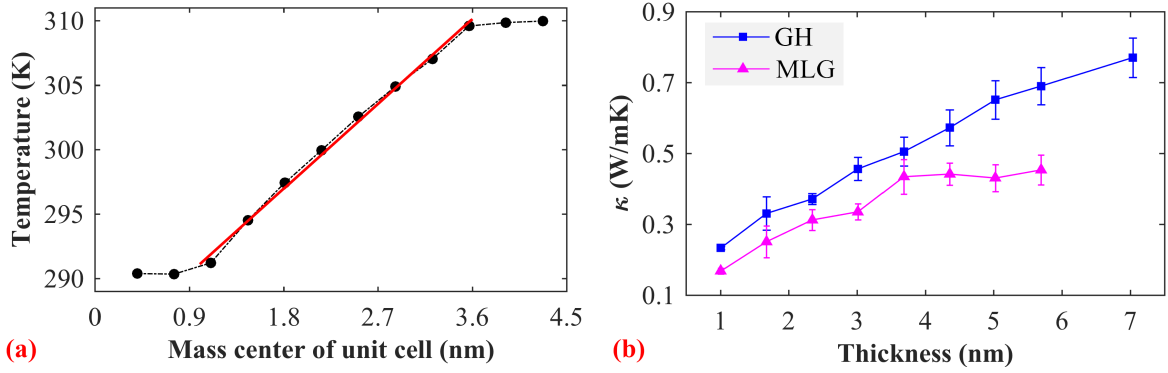

**Figure 2**. Thermal conductivity calculation. (a) The temperature profile of GH. The dotted line with circle marker is MD results, and the red solid line is the linear fitting. (b) The thermal conductivity of GH and MLG as a function of the sample thickness $L$.

The distinctly different size dependence reveals different thermal transport mechanism in GH with respect to that in MLG. To exploit it, in **Figure 3a,** we plot the thermal conductivity versus thickness curves in log-log scale. As shown, the thermal conductivity of GH can be well described by $\kappa \sim L^{0.62}$ for the examined sample length. Such observations are further validated from the relationship between the stationary time-averaged heat current and sample thickness (see Supporting Information **S2**). This power law dependent thermal conductivity has been predicted theoretically and confirmed experimentally in other quasi-one-dimensional nanomaterials, including carbon nanotubes,[33, 35-36] semiconductor nanowires,[37-38] polymers[39-40] and graphene nanoribbons.[41-49] It is clear now that in contrast with bulk materials, thermal conductivity of quasi-one-dimensional nano material diverges with the system size $L$ as $L^{\beta}$, because the momentum conservation in generic one-dimensional lattice will lead to super-diffusive heat transport[33, 50-53] and divergent thermal conductivity,[54-55] which is supported by rigorous mathematical proof.[56-58] Especially, thermal conductivity of graphene sheet does not saturate even in a large sample size of 9 μm.[49] The divergent nature of the thermal conductivity of graphene is also discussed earlier based on the framework of Klemens approximation.[59-60] To a certain degree, GH can be seen as curved graphene nanoribbon structure. The axial thermal conduction of GH is contributed by two parts: through interlayer vdW interaction and the in-plane thermal conduction of GNR projected to its axial direction. As the thermal conductivity of GNR diverges with length $L$ as $L^{\beta}$, the projected component, i.e., axial thermal conductivity of GH, also diverges with thickness. Therefore, for GH in the thickness of hundreds nanometers our results suggest that the thermal conductivity can be very high.

As aforementioned, the continuous nature of the GH structure makes both in-plane and out-of-plane phonon modes act as heat carrier. To further affirm this assumption, we compare the heat transfer characteristics of GH with that of graphene nanoribbon (GNR). Recall the atomic configurations in Figure 1c, we estimated the thermal conductivity of hydrogen terminated GNR with a size similar as that of the equivalent GNR projected along the axial direction of GH. A similar power-law relationship is found between $\kappa$ and length $L$ for hydrogen-passivated GNR (see Supporting Information **S3**), with the estimated exponent around 0.8 (i.e., $\kappa \propto L^{0.8}+C$). For a pristine GNR, the exponent is found to vary from 1 (for a narrow ribbon) to 3/2 (for a wider graphene sheet).[61] Similar It is worthy to mention that the exponent value of GNR is higher than that of the GH, which is in line with previous results that interlayer vdW interactions (as existed in GH structure) will suppress the size dependence of $\kappa$.[62]

Obviously, the thermal conductivity of MLG deviates from such power law dependence, and converges to a constant thermal conductivity when the thickness is larger than 4 nm, which signifies normal thermal conduction in MLG, and can be described by the kinetic theory. Following the kinetic theory,[63] the inverse of $\kappa$ follows a linear relationship with the inverse of thickness for MLG (Figure 3b), i.e., $1/\kappa = 1/\kappa_\infty(1+\lambda/L)$. Here, $\lambda$ and $\kappa_\infty$ are the cross-plane effective mean free path (MFP) in MLG structure and converged thermal conductivity of an infinitely thick MLG, respectively. As shown in Figure 3b, for MLG, $1/\kappa$ follows significant linear relationship with $1/L$, and it is a direct consequence of different nature of phonon transport compared with that for GH. It is worth mentioning that compared with the cross-plane thermal conductivity in few-layer graphene, the increase in thermal conductivity in graphene helicoid is limited. This can be understood from three facts. Firstly, the helicoid structure is made from graphene nanoribbon, which has much smaller thermal conductivity than that of infinite graphene sheet due to strong boundary phonon scattering. Secondly, the interlayer interactions are found to significantly suppress the in-plane thermal conductivity. Thirdly, for the stress-free helical structure, the axial thermal conductivity is partially contributed by the in-plane phonon transport projected to the axial direction, which is expected to hamper the "effective" transportation within the helicoid structure.

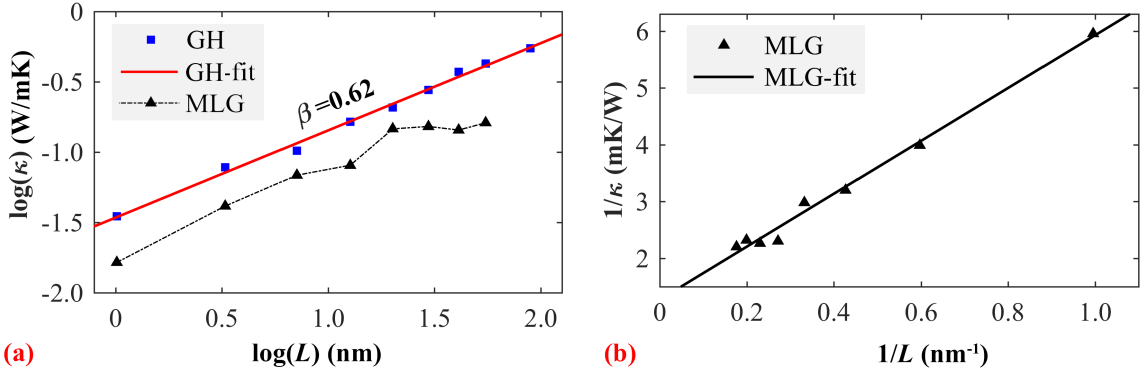

**Figure 3**. Thickness dependence of the thermal conductivity. (a) The logarithm relationship between $\kappa$ and thickness $L$; (b) The linear scaling relationship between the inverse of $\kappa$ and the inverse of $L$ for MLG structure.

**3.2 Better alignment results in high thermal conductivity**

In addition to the different size dependence, even in the short thickness range, the axial thermal conductivity of GH is still remarkably higher than the cross-plane thermal conductivity of MLG, signifying an additional underlying mechanism. Both experiments and simulations have shown that graphene has an ultra-low friction, or a superlubricity characteristic due to the incommensurability between adjacent layers.[64-65] The ultra-low interlayer friction feature makes the MLG structure very vulnerable to lateral perturbations, and leads to much higher interlayer misalignments. To quantitatively explore the misalignment, we estimate the average interlayer centroid misfit ($d_{cm}$) according to

$$d_{cm} = \frac{\sum_{i=4}^{N-3}(d_{i-1,i} + d_{i,i+1})}{N-8} \qquad (1)$$

Here $i$ ranges from 4 to $N$-3 to exclude both the fixed end and the heat sink/source, and $d$ is the centroid distance between two adjacent unit cells.

As illustrated in **Figure 4**a, the interlayer centroid misfit of MLG is fluctuating around 1.5 Å, which is nearly two times than that of the GH (~ 0.8 Å). This indicates that each layer in MLG structure has a larger mismatch with its adjacent upper or lower layers compared with that of GH, which substantially weaken the cross-layer heat transfer. Evidently, the smaller interlayer centroid misfit for GH is maintained due to its helicoid structure, which also endows the structure with a better stability for larger thickness (or higher layer number) compared with that of MLG. Furthermore, we calculate the overall axial misalignment ($d_{am}$) for GH and MLG, respectively, which reflects how much centroid of each unit is deviated from the ideal axis of the structure. Specifically, the overall axial misalignment is defined as $d_{am} = \max(\sum_{i=4}^{N-3} d_i)$, where $d_i$ is the vertical distance of the unit centroid with the ideal axis of GH/MLG. As shown in Figure 4b, the MLG has much larger overall axial misalignment and it increases significantly when the layer number increases. In comparison, the axial misalignment is much smaller for GH. This observation suggests that the alignment of MLG is more sensitive to lateral perturbations than that of the GH, especially with higher layer number, which weakens the cross-plane heat transfer. Similar observations of the impacts from the interlayer alignment on the thermal conductivity of laminate[66] and composites[67] with graphene have also been reported experimentally.

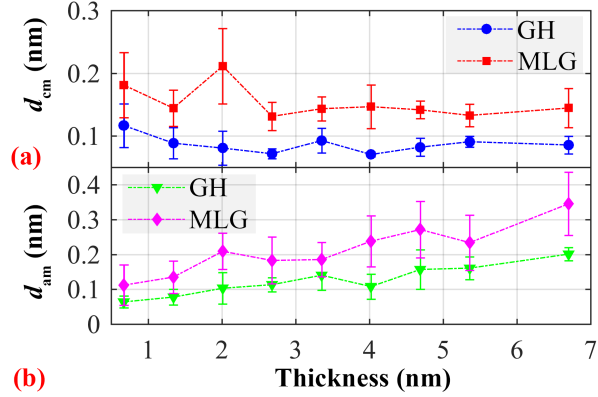

**Figure 4**. Interlayer misalignment of GH and MLG. (a) The average interlayer centroid misfit $d_{cm}$, and (b) the overall axial misalignment $d_{am}$.

**3.3 Strain effects on thermal conductivity of GH: significantly differ from MLG**

Above discussions have shown that the GH structure has superior thermal transport capability than that of MLG due to their structural difference. Earlier study shows that the Kapitza resistance in few-layer graphene (FLG) can be effectively controlled by applying cross-plane tensile strain.[68] However, given the fact that the interlayer interaction in graphene is constrained by weak vdW interactions, the effective applicable strain range is very small (about 10%). In this regard, the GH structure offers a facile way for thermal engineering under tensile strain, as its elastic deformation region can go up to a strain of 1000%.[19] As such, it is of great interest to assess how the thermal transport property of GH would change by applying axial strain. For such purpose, axial strain was firstly applied to both GH and MLG structures. The strained structures were firstly relaxed with fixed boundary condition for 200 ps and then heat baths were added to the system to calculate the thermal conductivity. Our previous work has shown that there are three distinct elastic deformation stages for the GH structure under tension, i.e., (1) an initial delamination stage for small strain, (2) a stable delamination, and (3) the stretch deformation of the fully delaminated structure.[19] We only focus on the heat transfer of the GH at the first and second stages of elastic deformation. It is found the strain energy of the GH keeps a constant with the relaxation time, indicating that the minimum energy status of the structure has been reached (see Supporting Information **S4**). For comparison, the influence from compressive stress will also be discussed, and the thermal conductivity of the strained MLG structure is also estimated for comparison purpose. The tensile and compressive deformation of GH and MLG are briefed in Supporting Information S4.

**Figure 5a** compares the $\kappa$ of GH and MLG as a function of the strain (within the initial delamination stage under tension less than 10% where the adjacent layers are still effectively constrained by vdW interactions). In general, the GH exhibits a decreasing $\kappa$ when the axial strain varies from compressive to tensile, which is consistent with the results from other carbon nanomaterials.[68-70] The MLG structure shows a decreasing $\kappa$ when the tensile strain incrases, but the compressive strain exerts ignorable influences when it is less than ~ 3% and reduces the thermal conductivity with further increment. Such observation can be explained from the perspective of interlayer shift under axial strain. As aforementioned, the GH structure has much better interlayer alignment. In contrast, the MLG interlayer is only constrained by vdW interactions, which makes the individual layer vulnerable to lateral displacement. Estimations show that the MLG has much higher average interlayer centroid misfit ($d_{cm}$) and overall axial misalignment ($d_{am}$) compared with that of the GH under compressive strain (see Figure 5b). When compressive strain increases, $d_{cm}$ increases abruptly for the MLG structure. Overall, for the MLG structure, although the enhanced interlayer interactions under compressive strain increases the phonon thermal conductivity, the enhancement is supposed to be cancelled out by increasing misalignment. Thus, an increasing $\kappa$ caused by compressive strain observed from GH structure is not seen in MLG structure.

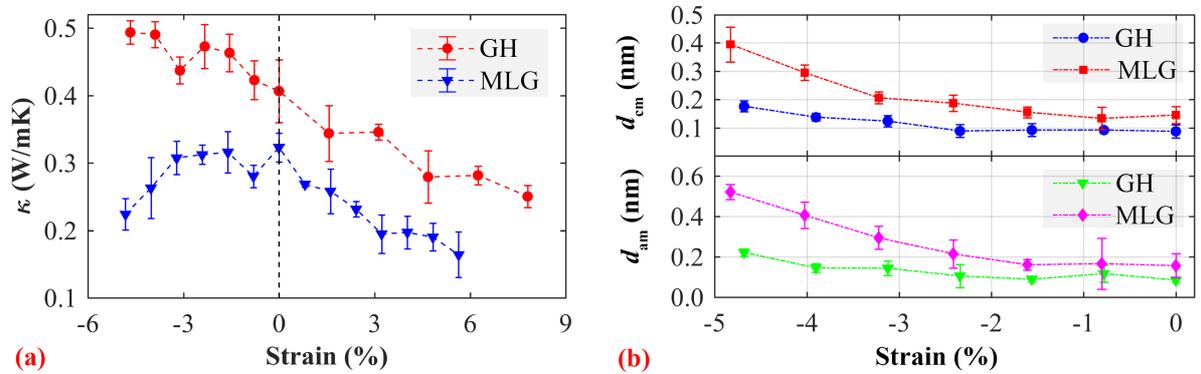

**Figure 5.** Thermal conductivity of GH/MLG under small axial/cross-plane strain. (a) Thermal conductivity of GH and MLG as a function of axial/cross-plane strain $\varepsilon$. (b) The average interlayer centroid misfit $d_{cm}$ (upper panel) and the overall axial misalignment $d_{am}$ (bottom panel) of GH and MLG under compressive strain.

The variations of the heat transfer between GH and MLG under axial strain can be further explained from the change of phonon population. Based on the vibrational density of states (VDOS), the phonon population variation is defined as the ratio between occupations of GH and MLG at the same strain status,[71]

$$\Delta n(\omega) = \frac{1+\int_0^\omega \text{VDOS}_{\text{GH},\varepsilon}(\omega')d\omega'}{1+\int_0^\omega \text{VDOS}_{\text{MLG},\varepsilon}(\omega')d\omega'} - 1 \quad (2)$$

Here, $\text{VDOS}_{\text{GH},\varepsilon}(\omega')$ and $\text{VDOS}_{\text{MLG},\varepsilon}(\omega')$ represent the VDOS at the strain of $\varepsilon$ for GH and MLG, respectively. VDOS is calculated from the Fourier transformation of the averaged velocity auto-correlation function (VACF). For both GH and MLG structures, we can clearly see the blue and red shift of the low-frequency phonon modes under compressive and tensile strain, respectively (see Supporting Information **S5**). **Figure 6** compares the in-plane (LA+TA+LO+TO) and out-of-plane (ZA+ZO) phonon occupation variation under compressive, zero, and tensile strain, respectivaly. As is seen, in all three scenarios, the GH structure uniformaly possesses much larger population of out-of-plane modes ($\Delta n > 0$). Such observation well aligns with the results in Figure 5a, affirming that GH has better thermal transport capability compared with MLG strcture. It is worth noting that the in-plane phonon modes also contribute to the heat transfer in GH structure, which are marginally affected by the axial strain compared with that of out-of-plane modes.

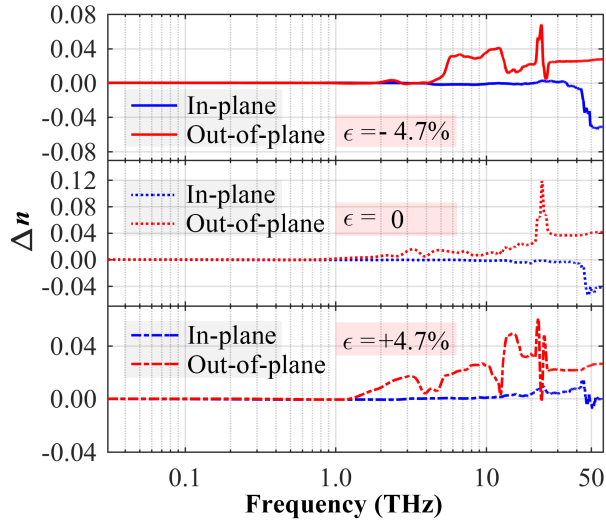

**Figure 6.** Phonon occupation variation of GH compared with MLG under cross-plane strain. Negative and positive strain denotes compressive and tensile strain, respectively.

After the initial delamination stage, the GH structure undergoes stable delamination process, while the MLG structure will separate and lose the heat transfer capability (due to the vanishing of interlayer vdW interactions). The GH structure will have two distinct sections (see **Figure 7**a). One is the delaminated region, which is like a curved nanoribbon (without interlayer vdW interactions). The other one is the un-delaminated region where the vdW interactions are still retained. We note that such inhomogeneous deformation scenario in GH structure is due to the fact that the partially delaminated structure is energetic favourable than homogeneously full-delaminated structure.[19] As illustrated in Figure 7a, the transition between un-delaminated and delaminated regions generates a clear temperature drop (see Supporting Information **S6** for more temperature profiles at different strain values). Considering the existence of the interfacial Kaptiza resistance and also the inhomogeneous structure, it is hard to arrive an accurate definition of thermal conductivity. As such, we compare the heat current in the GH structure with varying tensile strain, with the fixed temperature difference $\Delta T = 20$ K between the two heat baths. According to Figure 7b, the heat current ($Q$) experiences a relatively large reduction at the beginning of the deformation ($\varepsilon < \sim$ 100%), and then converges to around $0.5 \times 10^{-8}$ W. In general, a tensile strain can lead to an obvious reduction in the thermal conductivity of solid-state materials because of the increased anharmonicity and reduced phonon group velocity.[71-72] These results indicate that the GH structure maintains a stable heat transfer capability at large tensile strain. This is a consequence of its unique topological feature. In the delaminated region, as the angle between the thermal conductive sheet and the GH axis decreases, projected heat flux to axial thermal conductivity increases. Furthermore, in the un-delaminated region, the remained vdW force also supports a high heat conduction. Both of them benefit heat conduction, and result in the anomalous strain effect in GH in the large strain region.

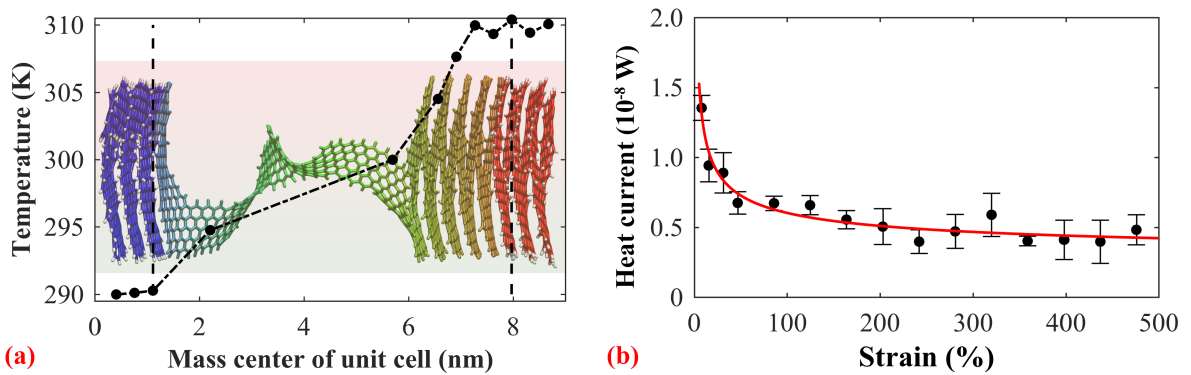

**Figure 7**. (a) Temperature profile at the strain of GH at the tensile strain of 85.84% (*x*-axis is the mass centre of each unit cell along the height direction). Inset shows the corresponding atomic configuration of the GH. (b) The heat current of GH as a function of tensile strain.

## 4. Conclusions

In summary, the helicoid structure offers a novel way to enhance the thermal transfer capability of multilayer graphene. Due to the interlayer misalignment, the inverse thermal conductivity of MLG exhibits a linear scaling relationship with the inverse of its thickness, and results in converged cross-plane thermal conductvity. In contrary, the continuous nature of GH not only endows the structure with better stability and alignment along the thickness direction, but also provides extra avenues for in-plane and out-of-plane phonon transfer in the axial direction. As a result, the thermal conductivity of GH shows a power-law relationship with its thickness, similar as that of graphene nanoribbon, and provides possibility to achieve high thermal conductivity. Moreover, unlike the MLG, the GH can withstand huge elastic deformation with tensile strain up to 1000%. This feature makes GH as a promising candidate for thermal engineering under cross-plane strain. It is found that the heat current experiences a relatively large reduction when the tensile strain is less than around 100%, and then converges to a constant value afterwards.

**Supporting Information**

The Supporting Information is available free of charge, including: additional results for the thermal conductivity of GH; logarithm relationship between heat current and smaple thickness; thermal conductivity of graphene nanoribbon; tensile and compressive deformation of GH and MLG; VDOS of GH and MLG under cross-plane strain; and temperatre profiles of GH at different tensile strain.

**AUTHOR INFORMATION**

**Corresponding Author**

*E-mail: zhangg@ihpc.a-star.edu.sg; yuantong.gu@qut.edu.au

**Author Contributions**

H.Z. carried out the simulation. H.Z., G.Z. C.Y., and Y.G. conducted the analysis and discussion.

**Notes**

The authors declare no competing financial interests.

**ACKNOWLEDGEMENT**

Supports from the ARC Discovery Project (DP170102861) and the High Performance Computing (HPC) resources provided by the Queensland University of Technology (QUT) are gratefully acknowledged. This research was undertaken with the assistance of resource and services from Intersect Australia Ltd, and the National Computational Infrastructure (NCI), which is supported by Australian Government. H.Z. would also like to acknowledge the support from the Start-up Fund from Western Sydney University.

**Table of Contents Graphic**

Graphene helicoid possesses the same amount of carbon atoms as its multilayer counterpart, while exhibiting distinct thermal transport properties with a different thickness dependency and strain dependency.

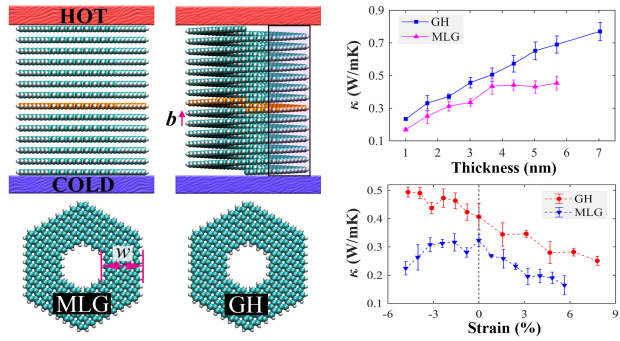